\documentclass[twocolumn,amsmath,amssymb,nobibnotes,aps,pra,superscriptaddress,reprint]{revtex4-2}

\usepackage{microtype,comment}
\usepackage{mathtools,amsmath}
\usepackage{array}
\usepackage{lipsum}
\usepackage{bbm}
\usepackage{amsthm,mathrsfs}
\usepackage{tikz-cd}
\usepackage{tikz}
\usetikzlibrary{arrows.meta, decorations.pathreplacing}

\usepackage{hyperref}

\usepackage{amsmath,amssymb,amsfonts,bbm,graphicx,hyperref,color,times}
\usepackage{subfigure}
\usepackage[english]{babel}

\newcommand{\bra}[1]{\ensuremath{\langle #1|}}
\newcommand{\ket}[1]{\ensuremath{|#1 \rangle}}
\newcommand{\braket}[2]{\ensuremath{\langle #1| #2 \rangle}}
\newcommand{\ketbra}[2]{\ensuremath{| #1 \rangle\hspace{-2pt} \langle #2 |}}

\definecolor{ncblue}{RGB}{200,220,255}
\definecolor{qred}{RGB}{255,210,210}

\begin{document}

\author{A. Navoni}
\affiliation{Universit{\`a} degli Studi di Milano, Dipartimento di Fisica, Via Celoria 16, I-20133 Milano, Italy}
\affiliation{ICFO – Institut de Ciencies Fotoniques, The Barcelona Institute of Science and Technology, Castelldefels, Barcelona 08860, Spain}
\author{M.G. Genoni}
\affiliation{Universit{\`a} degli Studi di Milano, Dipartimento di Fisica, Via Celoria 16, I-20133 Milano, Italy}
\affiliation{Istituto Nazionale di Fisica Nucleare, Sezione di Milano, via Celoria 16, 20133 Milan, Italy}
\author{A. Smirne}
\affiliation{Universit{\`a} degli Studi di Milano, Dipartimento di Fisica, Via Celoria 16, I-20133 Milano, Italy}
\affiliation{Istituto Nazionale di Fisica Nucleare, Sezione di Milano, via Celoria 16, 20133 Milan, Italy}

\title{Quantum contextuality from measurement invasiveness}
\begin{abstract}
Contextuality is a defining feature that separates the quantum from the classical descriptions of physical systems. 
Within the marginal-scenario framework, noncontextual models are characterized by the existence of a single joint probability distribution consistent with all measurable contexts, while contextual models violate this condition.
Building on this approach, we introduce a general method to analyze contextuality in terms of stochastic linear maps that
effectively model invasive measurements on an otherwise classical statistics.
These maps transform probabilities within the noncontextuality polytope, which includes all classical probabilities, into probabilities 
that may lie outside the polytope, while preserving the compatibility structure of the scenario at hand.
We derive general consistency conditions that such maps must satisfy to represent admissible invasive measurements, 
and we fully identify them for a paradigmatic example of contextuality for a single three-level quantum system. 
Furthermore, we introduce a quantifier of contextuality based on the minimal invasiveness required to reproduce a given probability distribution,
which offers a distinct approach on how to evaluate the degree of contextuality
in a general scenario.
\end{abstract}

\maketitle

\section{Introduction}
Quantum mechanics is a contextual theory, 
meaning that the value of an observable
cannot be assigned independently of the set of compatible observables being measured with it. 
Firstly identified by the Kochen-Specker theorem \cite{Kochen1967} 
as one of the key properties distinguishing quantum mechanics from any classical model,
contextuality is now recognized as a cornerstone of the foundations of quantum mechanics \cite{Budroni2022},
and has become a subject of growing interest across different fields, especially 
due to its emerging important role in understanding the origin of quantum computational advantage~\cite{Raussendorf2013,Howard2014,Vega2017,Bravyi2020}. 

The diverse approaches to contextuality \cite{Spekkens2005,Khrennikov2009,Abramsky2011,Kurzynski2012,Chaves2012,Cabello2014,Dzhafarov2017,Muller2023} 
reflect the multifaceted nature of the concept. 
Unlike the original formulation by Kochen and Specker, the more recent perspectives are based on theory-independent definitions of (non)contextuality, 
enabling a comparison between classical and quantum theories on neutral ground.
Among these, the marginal-scenario framework \cite{Chaves2012,Fritz2013} focuses directly on the probability distributions 
associated with joint measurements of compatible observables, imposing consistency constraints among different measurements 
and defining noncontextuality as the existence of a single global joint probability distribution consistent with all measurements.
Analogous to how Bell inequalities delineate the boundaries of local models \cite{Brunner2014},
noncontextual models are identified by specific sets of inequalities \cite{Budroni2022}, 
nonlocality being in fact a special instance of contextuality \cite{Fine1982,Abramsky2011}.
These inequalities define the facets of a polytope, 
which provides a clear and general geometric understanding of the separation between classical and quantum probabilities.

In this paper, we develop further this picture  by introducing linear maps, 
which transform classical probabilities within the noncontextuality polytope into quantum probabilities possibly lying outside it.
These maps are interpreted as modeling invasive measurements 
that stochastically modify the assigned values of classical observables,
and they allow for a systematic characterization of contextual probabilities via mathematical objects 
that are defined directly on the noncontextuality polytope.

We first derive the general consistency conditions that the stochastic maps must satisfy to preserve the compatibility structure of the model at hand,
and we explicitly work them out for the Klyachko-Can-Binicio{\v g}lu-Shumovsky (KCBS) scenario \cite{Klyachko2008},
a paradigmatic setting that proves contextuality of quantum mechanics even for a single three-level system.
Furthermore, we introduce a general quantifier of contextuality
that captures the cost of the invasiveness required to reproduce given quantum probabilities: 
after constructively illustrating its significance using quantum probabilities obtained from an experiment \cite{Lapkiewicz2011}, 
we evaluate it across a diverse set of quantum states in the KCBS scenario.

We stress that our approach is fully general and can be applied to arbitrary marginal scenarios. Indeed, the specific traits of contextuality in each scenario are reflected in the corresponding properties of the maps acting on the associated noncontextuality polytope.
Although these maps are represented by stochastic matrices, i.e., the same mathematical objects that describe transformations of classical probability distributions, they should not be interpreted as representative of classical physical processes. For instance, 
in Bell scenarios they will be associated also with nonlocal influences of each measurement apparatus on the distant outcome statistics, without implying any commitment to 
the ontological status of such influences, which falls outside the scope of our investigation. Our aim is instead to provide a scenario-independent 
framework that quantitatively connects contextuality and invasiveness through simple mathematical objects defined directly on the noncontextuality polytope.

The remainder of the paper is organized as follows. In Sec.\ref{sec:msa}, we fix the notation by briefly recalling
the marginal-scenario approach to contextuality. In Sec.\ref{sec:imm}, we introduce the main tool of our approach, i.e., 
the invasive-measurement maps, along with their general properties and connection with the noncontextuality polytope.
In Sec.\ref{sec:etk}, we treat the KCBS scenario: after providing an explicit description the KCBS noncontextuality polytope directly in terms
of the probability distributions -- indeed, equivalent to the one in terms of the correlations \cite{Araujo2013} -- 
we fully characterize the invasive-measurement maps for this scenario.
In Sec.\ref{sec:qct}, we introduce the quantifier of contextuality based on invasive measurements,
and explicitly evaluate it for the KCBS scenario, further comparing it 
with quantifiers that directly assess the distance of quantum probabilities from the noncontextuality polytope.
Finally, a summary of the results of the paper and future perspectives are given in Sec.\ref{sec:con}.

\section{Marginal scenarios and definition of contextuality}\label{sec:msa}
We start by recalling the marginal-scenario framework for contextuality \cite{Chaves2012,Fritz2013},
and by introducing a compact notation that will be useful in the following.

Given a set of observables
$\mathbbm{G}=\left\{A_1,\ldots, A_g \right\}$ with values $a_1, \ldots, a_g$ in the finite sets $O_1$, \ldots $O_g$,
a marginal scenario is a collection $\mathfrak{F} = \left\{G_1, \ldots, G_f\right\}$ of subsets of $\mathbbm{G}$,
$G_i \subset \mathbbm{G}$, such that if $G\in\mathfrak{F}$
and $G'\subset G$ then $G'\in\mathfrak{F}$. Each subset $G_i$ 
is a context and the description of a given marginal scenario within a theory consists in the definition of joint probability distributions $P^{G_i}$
of the quantities belonging to the same context $G_i$.
If some observables belong to more than one context, they
can be consistently recovered via marginalization from each of the contexts they belong to:
\begin{equation}\label{eq:marsccon}
P^{G_i}|_{G_i\cap G_j} = P^{G_j}|_{G_i\cap G_j},
\end{equation}
where $|_{G_i\cap G_j}$ denotes the restriction of the joint probability distribution to the observables that belong to both contexts $G_i$ and $G_j$,
obtained by marginalizing over the outcomes of the observables that do not belong to $G_i\cap G_j$.
The implication that if $G\in\mathfrak{F}$
and $G'\subset G$ then $G'\in\mathfrak{F}$, along with Eq.(\ref{eq:marsccon}), implies
that all observables within a context $G_i$ are compatible observables
with joint measurement described by $P^{G_i}$.

As an example, take the observables
$\mathbbm{G}=\left\{A_1, A_2, A_3\right\}$ and the marginal scenario described by 
\begin{equation}\label{eq:fex}
\mathfrak{F}= \left\{\left\{A_1\right\},\left\{A_2\right\},\left\{A_3 \right\},
\left\{A_1,A_2\right\},\left\{A_2, A_3\right\}\right\}.
\end{equation}
A description of such a scenario contains the probability distributions
(using the short-hand notation $P^{A_i}$ for $P^{\left\{A_i\right\}}$ and $P^{A_{i,j}}$
 for $P^{\left\{A_i, A_j\right\}}$)
 $P^{A_i}$, $i=1,2,3$, as well as the joint probability distributions $P^{A_{1,2}}$ and $P^{A_{2,3}}$,
 such that the following compatibility conditions hold:
 \begin{eqnarray}
&&\sum_{a_2}P^{A_{1,2}}(a_1,a_2) = P^{A_{1}}(a_1) \nonumber\\
&& \sum_{a_2}P^{A_{2,3}}(a_2,a_3) = P^{A_{3}}_S(a_3) \nonumber \\
&&\sum_{a_1}P^{A_{1,2}}(a_1,a_2) = \sum_{a_3}P^{A_{2,3}}(a_2,a_3) 
= P^{A_{2}}(a_2). \label{eq:comex}
\end{eqnarray}
The marginalization conditions above express in a consistent way that $\left\{A_1, A_2\right\}$ and $\left\{A_2, A_3\right\}$
are the joint measurements of, respectively, $A_1$ and $A_2$, and $A_2$ and $A_3$.

A collection of probability distributions $\left\{P^{G_i}\right\}_{G_i \in \mathfrak{F}}$ 
satisfying Eq.(\ref{eq:marsccon}) is called a marginal model of $\mathfrak{F}$.
A marginal model is defined as noncontextual when there is a joint probability distribution
$P^{joint}(a_1,\ldots,a_g)$ for all observables in $\mathbbm{G}$,
from which the probability distributions of the model are recovered via marginalization:
\begin{equation}\label{eq:defnc}
P^{G_i} = P^{joint}|_{G_i} \quad \forall G_i \in \mathfrak{F}.
\end{equation}
Indeed, in the scenario given by Eq.(\ref{eq:fex}), this corresponds to the existence of a joint probability distribution $P^{joint}(a_1, a_2, a_3)$
such that 
 \begin{eqnarray}
&&P^{A_{1,2}}(a_1,a_2) = \sum_{a_3} P^{joint}(a_1, a_2, a_3) \nonumber\\
&& P^{A_{2,3}}(a_2,a_3) = \sum_{a_1} P^{joint}(a_1, a_2, a_3); \label{eq:ncex}
\end{eqnarray}
note that the single-observable probability distributions can be obtained from $P^{joint}(a_1, a_2, a_3)$
via the compatibility relations already contained in Eq.(\ref{eq:comex}).

All classical models -- i.e., the observables in $\mathbbm{G}$ are random variables on a common phase space
and the states are probability distributions therein--
are noncontextual, and any noncontextual model is equivalent to a classical model -- with phase space 
$\Omega=O_1 \times \ldots \times O_g$, state $P^{joint}$ and random variables given by the Cartesian projections
from $\Omega$ to each $O_i$; thus, we will use as synonyms the words classical and noncontextual referred to models or probabilities.
On the other hand, quantum mechanics provides us with contextual models; the key example is represented by the Bell scenario,
where each $G_i$ is made of observables referred to different parties, Eq.(\ref{eq:marsccon}) corresponds to the non-signaling condition and the violation of Bell inequalities
is equivalent to the non-existence of a joint probability such that Eq.(\ref{eq:defnc}) holds \cite{Fine1982,Brunner2014}.

Given a model $\left\{P^{G_i}\right\}_{G_i \in \mathfrak{F}}$, we denote as $\vec{P}$ the $n$-dimensional vector
made of the probabilities of the contexts, where $n = \sum_{i=1}^{f} m_i$, with 
$m_i$ the number of probabilities in the $i-th$ context, i.e., the number
of outcomes for the measurement of the compatible observables in $G_i$;
the elements of $\vec{P}$ are ordered following the contexts' order, 
so that $P_1$ is associated with the first and $P_{m_1}$ with the last outcome of the joint measurement of the first context, $P_{m_1+1}$ is associated with the first of the second context and so on. Moreover,
we keep the letter $\vec{P}$ to denote probabilities of a generic theory, while we use
$\vec{C}$ ($\vec{Q}$) for probabilities of a classical (quantum) model
-- quantum probabilities are those that can be written as $\mbox{Tr}\left\{\rho \widehat{E}\right\}$,
where $\rho$ is a statistical operator and $\widehat{E}$ is an effect of a POVM \cite{Holevo2011,Heinosaari2014}.
Normalization of the probability distributions $\left\{P^{G_i}\right\}_{G_i \in \mathfrak{F}}$ means that
$$
\sum_{i=1}^{m_1} P_i = \ldots = \sum_{i=m_1+\ldots+m_{f-1}+1}^{m_1+\ldots+m_{f}} P_i = 1.
$$
This equality and the constraints in Eq.(\ref{eq:marsccon}) imply that not all probabilities $P_i$ are independent;
we denote as $\ell$ the number of independent probabilities and as $\vec{p}$ an $\ell$-dimensional vector of independent
probabilities. Normalization and the constraints in Eq.(\ref{eq:marsccon}) can be in fact expressed via  
an $n \times \ell$ matrix 
$\mathcal{M}$ and an $n$ dimensional vector $\vec{V}$ such that
\begin{equation}
\vec{P} = \mathcal{M} \vec{p} + \vec{V}. \label{eq:mvmv}
\end{equation}
Conversely, the vector with independent probabilities is obtained from the overall one via an $\ell \times n$ matrix $\mathcal{T}$,
\begin{equation}
\vec{p} = \mathcal{T} \vec{P}, \label{eq:tt}
\end{equation}
which singles out the independent components of the overall vector, i.e., it 
is a Cartesian projection from $\mathbbm{R}^{n}$ to $\mathbbm{R}^{\ell}$. Note that 
the choice of the couple $(\mathcal{M}, \vec{V})$, as well as of $\mathcal{T}$, for a given scenario $\mathfrak{F}$ is not unique 
due to the different choices of $\ell$ independent probabilities from all $n$ probabilities; 
from here on we will always imply that, given a scenario $\mathfrak{F}$, one fixed specific choice of $(\mathcal{M}, \vec{V})$
and $\mathcal{T}$ has been made.

\subsubsection*{The noncontextuality polytope}

For each marginal scenario $\mathfrak{F}$, the set of probability vectors arising 
from the noncontextual models is a convex set, 
that is, the polytope $\mathcal{N}_{\mathfrak{F}}$ spanned by the probability vectors $\vec{e}_\alpha$ 
obtained via Eqs.(\ref{eq:defnc}) and (\ref{eq:tt})
from the deterministic assignments of the observables' values in $\mathbbm{G}$ \cite{Budroni2022}; such a polytope is known as noncontextuality polytope. 

Being $\kappa_i$ the number of elements of $O_i$, there are then $\kappa= \kappa_1 \cdot \kappa_2 \cdot \cdot \cdot \kappa_g$ extremal points $\vec{e}_\alpha$ of the polytope,
and asking whether a given marginal model $\vec{P}$ is noncontextual corresponds to asking whether 
the vector of its independent probabilities $\vec{p}$ belongs to the polytope $\mathcal{N}_{\mathfrak{F}}$, i.e., can be written as
\begin{equation}\label{eq:polex}
\vec{p} = \sum_{\alpha=1}^{\kappa} \lambda_\alpha \vec{e}_\alpha, \qquad \lambda_\alpha \geq 0, 
\,\, \sum_{\alpha=1}^{\kappa} \lambda_\alpha = 1.
\end{equation}
Equivalently, each polytope can be defined as the intersection of a finite set of closed half-spaces, corresponding to its facets;
explicitly, for each scenario $\mathfrak{F}$, there is a
family of vectors $\vec{f}_i \in \mathbbm{R}^{\ell}$ and scalars $b_i \in \mathbbm{R}$ such that
\begin{equation}\label{eq:polyt}
\mathcal{N}_{\mathfrak{F}} = \{ \vec{p} \in \mathbb{R}^\ell \mid \vec{p} \cdot \vec{f}_i \leq b_i, \, i = 1, \ldots, r \}.
\end{equation} 
The enumeration of the facets of a noncontextuality polytope, given its extremal points, is in general an NP-hard problem \cite{Avis2004,Brunner2014}. 
The inequalities defining the nontrivial facets of $\mathcal{N}_{\mathfrak{F}}$ in Eq.(\ref{eq:polyt}) -- the trivial facets are those ensuring that
the vectors are made of positive elements and are thus satisfied by any theory -- are called
noncontextual inequalities, and they can be understood as the generalization of Bell's inequalities to generic marginal scenarios:
those quantum probabilities that are contextual
violate at least one of them.

\section{Invasive-measurement maps}\label{sec:imm}
The central idea of our approach is to determine the extent to which quantum probabilities outside the noncontextuality polytope 
can be described through some invasive measurement, consisting in a stochastic modification of an underlying classical statistics 
whose probabilities lie within the polytope.

Explicitly, given a vector of quantum probabilities $\vec{Q}$, its description
via an invasive measurement is given by a matrix $\mathcal{W}$ acting on a vector of classical probabilities $\vec{C}$,
$\vec{Q} = \mathcal{W} \vec{C}$.
In particular, we ask that $\mathcal{W}$ has a block-diagonal structure, so that it does not mix probabilities of different contexts, i.e., it has $f$ blocks on the diagonal, each with a $m_i \times m_i$ matrix $\mathcal{W}^{(i)}$,
\begin{equation}\label{eq:block}
\mathcal{W} = \bigoplus_{i=1}^{f} \mathcal{W}^{(i)}.
\end{equation}
Furthermore, we ask that 
each block $\mathcal{W}^{(i)}$ is a stochastic matrix, i.e., its elements $\mathcal{W}^{(i)}_{j,k}$ are positive and each column sum up to one:
\begin{eqnarray}
\mathcal{W}^{(i)}_{j,k} \geq 0, \qquad 
\sum_{j=1}^{m_i} \mathcal{W}^{(i)}_{j,k} =1. \label{eq:normw}
\end{eqnarray}
This is the key property for the interpretation of $\mathcal{W}$: the element $\mathcal{W}^{(i)}_{j,k}$ describes the conditional probability that,
performing a joint measurement of the $i$-th context,
if $k$ labels 
the underlying classical value
-- which would be the outcome of an ideal non-invasive measurement --
the actual measurement provides the outcome labeled by $j$ \cite{Richter2024}; 
any $\mathcal{W}^{(i)}_{j,k} \neq \delta_{j,k}$
can then be associated with some invasiveness in the measurement procedure.
We call \emph{invasive-measurement map} (IMM) any linear map from $\mathbbm{R}^n$ to $\mathbbm{R}^n$ satisfying Eqs.(\ref{eq:block}) and (\ref{eq:normw}).

The last requirement defining our approach is that
the maps accounting for invasiveness must preserve the consistency conditions in Eq.(\ref{eq:marsccon}), along with normalization.
In other terms, invasive maps must preserve the compatibility structure of the given scenario,
so that each of them represents
a consistent description of invasiveness for any set of classical probabilities of that scenario.
Recalling Eqs.(\ref{eq:mvmv}) and (\ref{eq:tt}), such a consistency requirement 
can be formulated through the commutativity of the following diagram:
\begin{equation}\label{eq:diag}
\begin{tikzcd}
\vec{C} \arrow[r, "\mathcal{W}"] \arrow[d, "\mathcal{T}"]
&[2em] \vec{Q} \arrow[d, shift left=1.5ex, "\mathcal{T}"] \\[2em]
\vec{c}  \arrow[r, "{(\mathcal{Z}, \vec{v})}"]  \arrow[u,shift left=1.5ex, "{(\mathcal{M}, \vec{V})}"]
& \vec{q} \arrow[u, "{(\mathcal{M}, \vec{V})}"]
\end{tikzcd},
\end{equation}
where $(\mathcal{Z}, \vec{v})$ denotes an affine map on the set of independent probabilities:
\begin{equation}\label{eq:qzv}
 \vec{q} = \mathcal{Z} \vec{c} +  \vec{v}.
\end{equation}
Commutativity at the "lower level" of the diagram,
$$
\mathcal{Z} \vec{c} + \vec{v} = \mathcal{T} \mathcal{W}\left(\mathcal{M}\vec{c}+\vec{V}\right) \quad \forall \vec{c},
$$
defines $(\mathcal{Z}, \vec{v})$ as the projection of $\mathcal{W}$ on the set of independent probabilities via $\mathcal{T}$, according to
\begin{equation}
 \mathcal{Z} =  \mathcal{T}  \mathcal{W} \mathcal{M}, \qquad
 \vec{v} = \mathcal{T}  \mathcal{W} \vec{V},\label{eq:constred}
\end{equation}
and therefore it holds for any given linear map $\mathcal{W}$.
On the other hand, $\mathcal{W}$ preserves normalization and the compatibility relations in Eq.(\ref{eq:marsccon})
if and only if commutativity holds at the "upper level" of the diagram, i.e.,
$$
\mathcal{M}\left(\mathcal{Z} \vec{c} + \vec{v}\right) + \vec{V} = \mathcal{W}\left(\mathcal{M}\vec{c}+\vec{V}\right) \quad \forall \vec{c},
$$ 
which is equivalent
to 
\begin{equation}\label{eq:mz}
\mathcal{M} \mathcal{Z} = \mathcal{W} \mathcal{M}, \quad \mathcal{M} \vec{v} + \vec{V} = \mathcal{W} \vec{V}.
\end{equation}
Replacing Eq.(\ref{eq:constred}) in Eq.(\ref{eq:mz}), we arrive at 
\begin{eqnarray}
 \mathcal{M}  \mathcal{T}  \mathcal{W} \mathcal{M}&=& \mathcal{W} \mathcal{M} \nonumber\\
\mathcal{M} \mathcal{T}  \mathcal{W} \vec{V}&=& \left(\mathcal{W}-\mathbbm{1}\right) \vec{V},\label{eq:const2}
\end{eqnarray}
which expresses the preservation of normalization and the compatibility structure of the scenario in terms of the map $\mathcal{W}$ only, and thus, along with Eqs.(\ref{eq:block}) and (\ref{eq:normw}), provides us with the constraints on the allowed $\mathcal{W}$; we call \emph{scenario-preserving} IMM any IMM satisfying Eq.(\ref{eq:const2}).

\begin{figure}[t]
    \centering
    \begin{tikzpicture}[scale=1.5, every node/.style={font=\footnotesize}]

  \coordinate (O) at (0,0);
  \def\radius{1.2}


  \coordinate (P1) at (0, 1.4);    
  \coordinate (P2) at (1.2, 0.15);  
  \coordinate (P3) at (0.6, -0.85); 
  \coordinate (P4) at (-0.8, -0.7); 
  \coordinate (P5) at (-0.9, 0.7);  

\fill[qred,opacity=0.8]
  (P1)
    .. controls +(0.9,0.3) and +(0.1,1.1) .. (P2)
    to[out=270, in=30, looseness=1.0] (P3)
    .. controls +(-0.5,-0.5) and +(0.1,-1.) .. (P4)
    -- (P5) -- cycle;
    
\fill[ncblue,opacity=0.9]
  (P1) -- (P2) -- (P3) -- (P4) -- (P5) -- cycle;  


 \draw[thick,dashed] (P1) -- (P2);                  
  \draw[thick, dashed] (P2) -- (P3);          
  \draw[thick,dashed] (P3) -- (P4);                  
  \draw[thick] (P4) -- (P5);
    \draw[thick] (P5) -- (P1);


  \coordinate (c) at (0.1,0.3);
  \filldraw[black] (c) circle (1pt) node[below right] {$\vec{c}$};

  \coordinate (q) at (0.83,1.1);
  \filldraw[black] (q) circle (1pt) node[below right] {$\vec{q}$};

 \draw[->, thick, black, shorten <=1pt, shorten >=1pt] 
  (c) -- (q) node[pos=0.35, above, sloped] {$(\mathcal{Z}, \vec{v})$};

\draw[thick, black]
(P1) .. controls +(0.9,0.3) and +(0.1,1.1) .. (P2);

\draw[thick, black]
 (P2) to[out=270, in=30, looseness=1.0] (P3);

\draw[thick, black]
  (P3) .. controls +(-0.5,-0.5) and +(0.1,-1.) .. (P4);

  \node at (-0.05,-0.15) {Noncontextual};

  \node at (-0.2,-1.0) {Quantum};

\end{tikzpicture} 
      \caption{Sketch of the action of the affine map $(\mathcal{Z}, \vec{v})$ obtained from an IMM $\mathcal{W}$ via Eq.(\ref{eq:constred}): 
  a vector of classical probabilities $\vec{c}$, within the noncontextuality polytope, is mapped to a vector of quantum probabilities $\vec{q}$, 
  possibly outside the polytope; the facets of the noncontextuality polytope are depicted as straight lines -- solid and dashed --
  while the border of the set of quantum probabilities as solid lines -- straight and curve; 
  the interior of the noncontextuality polytope is the light blue area,
  while the interior of the set of quantum probabilities is the light red and light blue areas. 
  The consistency conditions enforced by scenario preservation -- see the diagram (\ref{eq:diag}) -- set 
  the constraints in Eq.(\ref{eq:const2}).
}
  \label{fig:sch}
\end{figure}

Summarizing, given a quantum model of a scenario $\mathfrak{F}$ -- with the associated couple $(\mathcal{M}, \vec{V})$ and projection $\mathcal{T}$ -- its 
description
through invasive measurements on a classical model consists of a 
scenario-preserving IMM $\mathcal{W}$ 
such that 
the vector of independent probabilities $\vec{q}$ can be written as -- see Eq.(\ref{eq:polex}) --
\begin{equation}\label{eq:convexred}
\vec{q} = \sum_{\alpha=1}^\kappa \lambda_\alpha \mathcal{Z}\vec{e}_\alpha + \vec{v}, 
\qquad \lambda_\alpha \geq 0, 
\,\, \sum_{\alpha=1}^{\kappa} \lambda_\alpha = 1,
\end{equation}
where $(\mathcal{Z}, \vec{v})$ is the couple defined by Eq.(\ref{eq:constred}),
or, equivalently, if $\mathcal{Z}$ is invertible, such that -- see Eq.(\ref{eq:polyt}) --
\begin{eqnarray}\label{eq:continred2}
&&\vec{q} \cdot \vec{g}_i  \leq d_i, \quad i = 1, \ldots, r, \\
&&\vec{g}_i =  (\mathcal{Z}^{-1})^\top\vec{f}_i,\,\,\,\, d_i = b_i +  \vec{v} \cdot \vec{f}_i, \nonumber
\end{eqnarray}
where $^\top$ denotes transposition.

The action of the affine map $(\mathcal{Z}, \vec{v})$ connecting probabilities within the noncontextuality polytope
to quantum probabilities possibly outside it is sketched in Fig.\ref{fig:sch}.

\section{Example: the KCBS scenario}\label{sec:etk}
To illustrate our approach, we consider one specific representative marginal scenario,
that is the well-known KCBS scenario \cite{Klyachko2008}.
The latter concerns five observables,
$\mathbbm{G} = \left\{A_1,\ldots,A_5\right\}$ where each $A_i$ has outcomes $O_i=\left\{-1,1\right\}$, and it consists of
five contexts, each made of two observables and where each observable appears in two and only two contexts \footnote{We are
not taking into account explicitly the trivial contexts made of one single observable,
since the corresponding probabilities can always be recovered via marginalization from the probabilities
of the five contexts considered.}, according to
\begin{equation}\label{eq:KCBS}
\begin{split}
& \mathfrak{F}_{KCBS}=\{\left\{A_1,A_2\right\},\left\{A_2,A_3\right\},\left\{A_3,A_4\right\},\left\{A_4,A_5\right\},\left\{A_5,A_1\right\} \}.
\end{split}
\end{equation}
This is the simplest scenario leading to noncontextual inequalities that are violated by quantum mechanics,
as this happens for a Hilbert space of dimension 3.

\subsection{The KCBS polytope}
The  KCBS scenario is therefore defined by $f=5$ contexts and $m=4$ probabilities $P^{A_{i,i+1}}(a_i, a_{i+1})$ in each context, so that $\vec{C}$ and $\vec{Q}$ are $20$-dimensional vectors of probabilities.
The consistency conditions on marginalization in Eq.(\ref{eq:marsccon}) explicitly read 
(identifying here and in the following $i=6$ with $i=1$ and $i=0$ with $i=5$)
\begin{equation}\label{eq:markcbs}
    \sum_{a_{i+1}\pm 1}P^{A_{i,i+1}}(a_i,a_{i+1}) 
= \sum_{a_{i}=\pm 1}P^{A_{i-1,i}}(a_i,a_{i+1})
\end{equation}
and, along with the normalization conditions 
\begin{equation}\label{eq:norkcbs}
\sum_{a_i, a_{i+1}=\pm 1}P^{A_{i,i+1}}(a_i,a_{i+1}) = 1,
\end{equation}
represent 10 independent constraints
on the probabilities of the contexts -- 5 from normalization conditions within each context, and 5 from independent marginalization constraints, 
due to the fact that each observable is associated with two distinct contexts, so that each single-observable probability 
can be reconstructed equivalently from marginal distributions pertaining to two different contexts via equation (\ref{eq:markcbs});
we then have $\ell = n -10 = 10$ independent probabilities.

We order the possible outcomes of $\left\{A_i, A_{i+1}\right\}$ in each context
as $\left\{(+1,+1), (+1,-1), \right.$ $\left. (-1,+1), (-1,-1)\right\}$,
and choose the first two probabilities in each context as the independent ones -- i.e. those defining $\vec{p}$.
The projection $\mathcal{T}$ from $\mathbbm{R}^n$ to $\mathbbm{R}^{\ell}$ such that $\vec{p} = \mathcal{T} \vec{P}$
is then the $10 \times 20$ matrix whose rows pick the first two outcomes of each context, i.e., introducing its matrix elements $\mathcal{T}_{j,k}$,
\begin{eqnarray}
  &&\mathcal{T}_{1,1} = \mathcal{T}_{2,2} = \mathcal{T}_{3,5} = \mathcal{T}_{4,6} = \mathcal{T}_{5,9} =\mathcal{T}_{6,10} = 1 \nonumber\\
  &&\mathcal{T}_{7,13} =\mathcal{T}_{8,14}
 = \mathcal{T}_{9,17} =\mathcal{T}_{10,18} = 1 \nonumber\\
 &&\text{all other elements} \,\, = 0.  \label{eq:kcbst}
\end{eqnarray}
The marginalization and normalization constraints translate -- see Eq.(\ref{eq:mvmv}) --
into the 
$20$-dimensional vector
\begin{equation}\label{eq:kcbsv}
\vec{V} = \left(0, 0, 0, 1, 0, 0, 0, 1, 0, 0, 0, 1, 0, 0, 0, 1, 0, 0, 0, 1\right).     
\end{equation}
and the $20 \times 10$ matrix $\mathcal{M}$ composed of 5 identical blocks
\begin{equation}\label{eq:kcbsm}
\mathcal{M}^{(i)} = \left(
\begin{array}{cccc}
 1 & 0 & 0 & 0 \\
 0 & 1 & 0 & 0 \\
 -1 & 0 & 1 & 1\\
 0 & -1 & -1 & -1  \\
\end{array}
\right),
\end{equation}
with the top left element of $\mathcal{M}^{(i)}$ at row $4 i -3$ and column $2i-1$, the last two column of $\mathcal{M}^{(5)}$ being identified as (part of) the first two columns
of $\mathcal{M}$, and all other elements being equal to 0.

The noncontextuality polytope for $\mathfrak{F}_{KCBS}$, $\mathcal{N}_{KCBS}$, is given by the convex hull of the 32 extremal points
$\vec{e}_{\alpha}$
obtained from the projection of the probabilities vectors $\vec{P}$ corresponding to the deterministic assignements
of the dichotomic observables $A_i$, $i=1,\ldots, 5$. For example,
the assignment $(1,1,1,1,1)$ corresponds to
$$
\vec{P} = \left(1, 0, 0, 0, 1, 0, 0, 0, 1, 0, 0, 0, 1, 0, 0, 0, 1, 0, 0, 0\right)
$$
and then to
$$
\vec{e}_1 = \mathcal{T} \vec{P} =  \left(1, 0, 1, 0, 1, 0, 1, 0, 1, 0\right).
$$
The complete list of extremal points is 
\begin{eqnarray}
\vec{e}_1  &=& \left(1, 0, 1, 0, 1, 0, 1, 0, 1, 0\right) \nonumber\\
\vec{e}_2  &=& \left(1, 0, 1, 0, 1, 0, 0, 1, 0, 0\right) \nonumber\\
\vec{e}_3  &=& \left(1, 0, 1, 0, 0, 1, 0, 0, 1, 0\right) \nonumber\\
\vec{e}_4  &=& \left(1, 0, 1, 0, 0, 1, 0, 0, 0, 0\right) \nonumber\\
\vec{e}_5  &=& \left(1, 0, 0, 1, 0, 0, 1, 0, 1, 0\right) \nonumber\\
\vec{e}_6  &=& \left(1, 0, 0, 1, 0, 0, 0, 1, 0, 0\right) \nonumber\\
\vec{e}_7  &=& \left(1, 0, 0, 1, 0, 0, 0, 0, 1, 0\right) \nonumber\\
\vec{e}_8  &=& \left(1, 0, 0, 1, 0, 0, 0, 0, 0, 0\right) \nonumber\\
\vec{e}_9  &=& \left(0, 1, 0, 0, 1, 0, 1, 0, 1, 0\right) \nonumber\\
\vec{e}_{10} &=& \left(0, 1, 0, 0, 1, 0, 0, 1, 0, 0\right) \nonumber\\
\vec{e}_{11} &=& \left(0, 1, 0, 0, 0, 1, 0, 0, 1, 0\right) \nonumber\\
\vec{e}_{12} &=& \left(0, 1, 0, 0, 0, 1, 0, 0, 0, 0\right) \nonumber\\
\vec{e}_{13} &=& \left(0, 1, 0, 0, 0, 0, 1, 0, 1, 0\right) \nonumber\\
\vec{e}_{14} &=& \left(0, 1, 0, 0, 0, 0, 0, 1, 0, 0\right) \nonumber\\
\vec{e}_{15} &=& \left(0, 1, 0, 0, 0, 0, 0, 0, 1, 0\right) \nonumber\\
\vec{e}_{16} &=& \left(0, 1, 0, 0, 0, 0, 0, 0, 0, 0\right) \nonumber\\
\vec{e}_{17} &=& \left(0, 0, 1, 0, 1, 0, 1, 0, 0, 1\right) \nonumber\\
\vec{e}_{18} &=& \left(0, 0, 1, 0, 1, 0, 0, 1, 0, 0\right) \nonumber\\
\vec{e}_{19} &=& \left(0, 0, 1, 0, 0, 1, 0, 0, 0, 1\right) \nonumber\\
\vec{e}_{20} &=& \left(0, 0, 1, 0, 0, 1, 0, 0, 0, 0\right) \nonumber
\end{eqnarray}
\begin{eqnarray}
\vec{e}_{21} &=& \left(0, 0, 0, 1, 0, 0, 1, 0, 0, 1\right) \nonumber\\
\vec{e}_{22} &=& \left(0, 0, 0, 1, 0, 0, 0, 1, 0, 0\right) \nonumber\\
\vec{e}_{23} &=& \left(0, 0, 0, 1, 0, 0, 0, 0, 0, 1\right) \nonumber\\
\vec{e}_{24} &=& \left(0, 0, 0, 1, 0, 0, 0, 0, 0, 0\right) \nonumber\\
\vec{e}_{25} &=& \left(0, 0, 0, 0, 1, 0, 1, 0, 0, 1\right) \nonumber\\
\vec{e}_{26} &=& \left(0, 0, 0, 0, 1, 0, 0, 1, 0, 0\right) \nonumber\\
\vec{e}_{27} &=& \left(0, 0, 0, 0, 0, 1, 0, 0, 0, 1\right) \nonumber\\
\vec{e}_{28} &=& \left(0, 0, 0, 0, 0, 1, 0, 0, 0, 0\right) \nonumber\\
\vec{e}_{29} &=& \left(0, 0, 0, 0, 0, 0, 1, 0, 0, 1\right) \nonumber\\
\vec{e}_{30} &=& \left(0, 0, 0, 0, 0, 0, 0, 1, 0, 0\right) \nonumber\\
\vec{e}_{31} &=& \left(0, 0, 0, 0, 0, 0, 0, 0, 0, 1\right) \nonumber\\
\vec{e}_{32} &=& \left(0, 0, 0, 0, 0, 0, 0, 0, 0, 0\right). \label{eq:e32}\\
\nonumber
\end{eqnarray}

Equivalently, $\mathcal{N}_{KCBS}$ is defined by the inequalities in Eq.(\ref{eq:polyt})
for the following set of vectors and real numbers
\begin{eqnarray}
\vec{f}_1 &=& \left(0, 1, 0, 1, 0, 1, 0, 1, 0, 1\right), \quad b_1 = 2 \label{eq:kcbsineq16}\\
\vec{f}_2 &=& \left(-1, 0, 0, 1, 0, 1, 1, 0, 0, -1\right), \quad b_2 = 1 \nonumber\\
\vec{f}_3 &=& \left(-1, 0, 0, 1, 1, 0, -1, 0, 1, 0\right), \quad b_3 = 1 \nonumber\\
\vec{f}_4 &=& \left(0, 1, 0, 1, 1, 0, 0, -1, -1, 0\right), \quad b_4 = 1 \nonumber\\
\vec{f}_5 &=& \left(-1, 0, 1, 0, -1, 0, 0, 1, 1, 0\right), \quad b_5 = 1 \nonumber\\
\vec{f}_6 &=& \left(0, 1, 1, 0, -1, 0, 1, 0, -1, 0\right), \quad b_6 = 1 \nonumber\\
\vec{f}_7 &=& \left(0, 1, 1, 0, 0, -1, -1, 0, 0, 1\right), \quad b_7 = 1 \nonumber\\
\vec{f}_8 &=& \left(0, -1, -1, 0, 0, 1, 0, 1, 1, 0\right), \quad b_8 = 1 \nonumber\\
\vec{f}_9 &=& \left(1, 0, -1, 0, 0, 1, 1, 0, -1, 0\right), \quad b_9 = 1 \nonumber\\
\vec{f}_{10} &=& \left(1, 0, -1, 0, 1, 0, -1, 0, 0, 1\right), \quad b_{10} = 1 \nonumber\\
\vec{f}_{11} &=& \left(1, 0, 0, -1, -1, 0, 0, 1, 0, 1\right), \quad b_{11} = 1 \nonumber\\
\vec{f}_{12} &=& \left(-1, 0, 1, 0, 0, -1, 0, -1, 0, -1\right), \quad b_{12} = 0 \nonumber\\
\vec{f}_{13} &=& \left(0, -1, -1, 0, 1, 0, 0, -1, 0, -1\right), \quad b_{13} = 0 \nonumber\\
\vec{f}_{14} &=& \left(0, -1, 0, -1, -1, 0, 1, 0, 0, -1\right), \quad b_{14} = 0 \nonumber\\
\vec{f}_{15} &=& \left(0, -1, 0, -1, 0, -1, -1, 0, 1, 0\right), \quad b_{15} = 0 \nonumber\\
\vec{f}_{16} &=& \left(1, 0, 0, -1, 0, -1, 0, -1, -1, 0\right), \quad b_{16} = 0, \nonumber
\end{eqnarray}
which correspond to all its nontrivial facets \cite{Araujo2013}.

\subsubsection*{Quantum probabilities for the KCBS scenario in \texorpdfstring{$\mathbbm{C}^3$}{C3}}


The scenario $\mathfrak{F}_{KCBS}$ can be realized by quantum mechanics 
in $\mathcal{H} = \mathbbm{C}^3$ by means of the self-adjoint operators 
\begin{equation}\label{eq:sta}
\widehat{A}_i = 2 \ket{v_i}\bra{v_i}-\mathbbm{1}_{\mathbbm{C}^3},
\end{equation}
where
\begin{equation}\label{eq:vikcbs}
\begin{split}
&\ket{v_i} = \left(\cos\theta, \sin\theta \cos \varphi_i,  \sin\theta \sin\varphi_i\right),\\
&\cos\theta=\frac{1}{\sqrt[4]{5}},\\
& \varphi_1 =\frac{2 \pi}{5}, \varphi_2 =\frac{6 \pi}{5}, \varphi_3 =0, 
\varphi_4 =\frac{4 \pi}{5},  \varphi_5 =\frac{8 \pi}{5}.
\end{split}
\end{equation}
Since $\braket{v_i}{v_{i+1}} = 0$, the self-adjoint operators defined above are pairwise compatible, $[\widehat{A}_i, \widehat{A}_{i+1}]=0$.
Moreover, one can write each of them
as 
\begin{equation}
\widehat{A}_i = \ketbra{v_i}{v_i} - \ketbra{v_{i+1}}{v_{i+1}} - \ketbra{w_i}{w_i},  
\end{equation}
where $\ket{w_i}$ is the vector orthogonal to $\ket{v_i}$ and $\ket{v_{i+1}}$,
so that
\begin{equation}
\widehat{\Xi}_i =  \ketbra{v_i}{v_i}
\end{equation}
projects into the subspace $(+1,-1)$ for the joint measurement $\left\{\widehat{A}_i,\widehat{A}_{i+1}\right\}$ 
defining the $i$-th context, see Eq.(\ref{eq:KCBS}),
while the couple of outcomes $(+1,+1)$ occurs with probability zero.
Hence, recalling the choice of independent probabilities -- see before Eq.(\ref{eq:kcbst}) --
the vector of quantum probabilities $\vec{q}$ is fixed by
the vector $\vec{\widehat{\Pi}}$ of projectors
given by
\begin{equation}\label{eq:pix}
\vec{\widehat{\Pi}} = \left(\widehat{0},\widehat{\Xi}_1,\ldots,\widehat{0},\widehat{\Xi}_5\right),
\end{equation}
according to
\begin{equation}\label{eq:qx}
\vec{q}= \mbox{Tr}\left\{\vec{\widehat{\Pi}}\rho\right\}.
\end{equation}

Only the first of the inequalities defining the facets of $\mathcal{N}_{KCBS}$ -- see Eq.(\ref{eq:polyt}),
with $\vec{f}_1$ and $b_1$ given by Eq.(\ref{eq:kcbsineq16}) -- 
can be violated by quantum probabilities $\vec{q}$ arising from a generic state on $\mathcal{H}=\mathbbm{C}^3$ 
and the projective measurements defined by Eq.(\ref{eq:sta}); this inequality is in fact commonly referred to as \emph{the} KCBS inequality.
In particular, the state $\rho = \ket{1}\bra{1}$ leads to 
\begin{equation}\label{eq:max}
\sum_{i=1}^5 Q^{A_{i,i+1}}(a_i =+1, a_{i+i}=-1)\\
=\sum_{i=1}^5 \mbox{Tr}\{\widehat{\Xi}_i \rho\} = \sqrt{5} > 2,
\end{equation}
which represents a maximal violation of the KCBS inequality.

\subsection{Consistency constraints on the invasive-measurement maps}
Given the explicit characterization of the noncontextuality polytope for the KCBS scenario, 
we can now fully determine the corresponding scenario-preserving IMMs.
In fact, using Eqs.(\ref{eq:kcbst})-(\ref{eq:kcbsm}) the conditions in (\ref{eq:const2}) can be worked out explicitly, leading to
\begin{eqnarray}\label{eq: general constraints}
    \mathcal{W}_{11}^{(i)}+\mathcal{W}_{31}^{(i)}&=&\mathcal{W}_{13}^{(i)}+\mathcal{W}_{33}^{(i)} \nonumber\\
    \mathcal{W}_{11}^{(i)}+\mathcal{W}_{21}^{(i)}&=&\mathcal{W}_{12}^{(i)}+\mathcal{W}_{22}^{(i)} \nonumber\\
    \mathcal{W}_{12}^{(i)}+\mathcal{W}_{32}^{(i)}&=&\mathcal{W}_{14}^{(i)}+\mathcal{W}_{34}^{(i)} \nonumber\\
    \mathcal{W}_{13}^{(i)}+\mathcal{W}_{23}^{(i)}&=&\mathcal{W}_{14}^{(i)}+\mathcal{W}_{24}^{(i)} \nonumber\\
        \mathcal{W}_{11}^{(i)}+\mathcal{W}_{31}^{(i)}&=&\mathcal{W}_{11}^{(i+1)}+\mathcal{W}_{21}^{(i+1)} \nonumber\\
    \mathcal{W}_{12}^{(i)}+\mathcal{W}_{32}^{(i)}&=&\mathcal{W}_{13}^{(i+1)}+\mathcal{W}_{23}^{(i+1)}.
    \end{eqnarray}
Along with Eqs.(\ref{eq:block}) and (\ref{eq:normw}), these relations define the set of scenario-preserving IMMs
and thus they fix the set of invasive measurements that are consistent within our approach.

Interestingly, these constraints can be translated into some general structural properties
that characterize (consistent) measurement invasiveness accounting for contextuality in the KCBS scenario. First, if the measurement of the $i$-th context 
is non-invasive, i.e. $\mathcal{W}^{(i)}_{jk} = \delta_{j,k}$, 
it follows from Eqs.~(\ref{eq:normw}) and (\ref{eq: general constraints})
that the IMM elements are constrained by
\begin{eqnarray}
&& W_{11}^{(i-1)} = W_{22}^{(i-1)}, \quad W_{33}^{(i-1)}=W_{44}^{(i-1)}, \nonumber\\ 
&&W_{21}^{(i-1)}=W_{41}^{(i-1)}=W_{12}^{(i-1)}=W_{32}^{(i-1)} =W_{23}^{(i-1)}= 0 \nonumber \\
&&W_{43}^{(i-1)} =W_{14}^{(i-1)}=W_{34}^{(i-1)} = 0, \nonumber\\
&& W_{11}^{(i+1)}=W_{33}^{(i+1)}, \quad W_{22}^{(i+1)}=W_{44}^{(i+1)} \nonumber\\
&& W_{31}^{(i+1)}=W_{41}^{(i+1)}=W_{32}^{(i+1)}=W_{42}^{(i+1)} =W_{13}^{(i+1)} = 0 \nonumber\\
&&W_{23}^{(i+1)} =W_{14}^{(i+1)}=W_{24}^{(i+1)} = 0.\label{eq:imma}
\end{eqnarray}
From these, in turn it follows that an invasive measurement cannot occur both before and after non-invasive ones, i.e.,
\begin{equation}
\mathcal{W}^{(i+1)}_{jk} = \mathcal{W}^{(i-1)}_{jk} = \delta_{j,k} \,\, \forall \,\, j,k \,\,\Rightarrow \,\, \mathcal{W}^{(i)}_{jk} = \delta_{j,k}  \,\, \forall \,\, j,k ,
\end{equation}
which also means that invasive measurements must affect one contiguous block of contexts in $\mathfrak{F}_{KCBS}$.

Furthermore, we now show that in the KCBS scenario any probability vector $\vec{P}$ satisfying the normalization 
and marginalization conditions -- Eqs.(\ref{eq:markcbs}) and (\ref{eq:norkcbs}) -- can be obtained by applying a
scenario-preserving IMM $\mathcal{W}$ -- see Eqs.~(\ref{eq:block}), (\ref{eq:normw}) and (\ref{eq: general constraints}) --
to a classical probability vector $\vec{C}$; 
even more, such a classical probability vector can be taken as any vertex of the noncontextuality polytope.
The polytope of the probability vectors satisfying normalization 
and marginalization conditions, which is known as the non-disturbance polytope, has been fully characterized in \cite{Araujo2013}. 
Its facets are defined by the conditions ensuring the positivity of each element of $\vec{P}$, that is, in our language 
and denoting as $(\vec{J})_i$ the $i$-th element of the vector $\vec{J}$,
\begin{equation}
\left(\mathcal{M} \vec{p} + \vec{V}\right)_i \geq 0 \quad i=1,\ldots,20.
\end{equation}
The non-disturbance polytope has 48 extremal points, given by the 32 listed in Eq.(\ref{eq:e32}),
which are then common to the noncontextuality polytope $\mathcal{N}_{KCBS}$, and the further 16 vectors
\begin{eqnarray}
\vec{e}_{33} &=& \begin{pmatrix} 0 & \tfrac{1}{2} & \tfrac{1}{2} & 0 & \tfrac{1}{2} & 0 & \tfrac{1}{2} & 0 & \tfrac{1}{2} & 0 \end{pmatrix}, \nonumber \\
\vec{e}_{34} &=& \begin{pmatrix} \tfrac{1}{2} & 0 & 0 & \tfrac{1}{2} & \tfrac{1}{2} & 0 & \tfrac{1}{2} & 0 & \tfrac{1}{2} & 0 \end{pmatrix}, \nonumber \\
\vec{e}_{35} &=& \begin{pmatrix} \tfrac{1}{2} & 0 & \tfrac{1}{2} & 0 & 0 & \tfrac{1}{2} & \tfrac{1}{2} & 0 & \tfrac{1}{2} & 0 \end{pmatrix}, \nonumber \\
\vec{e}_{36} &=& \begin{pmatrix} \tfrac{1}{2} & 0 & \tfrac{1}{2} & 0 & \tfrac{1}{2} & 0 & 0 & \tfrac{1}{2} & \tfrac{1}{2} & 0 \end{pmatrix}, \nonumber \\
\vec{e}_{37} &=& \begin{pmatrix} \tfrac{1}{2} & 0 & \tfrac{1}{2} & 0 & \tfrac{1}{2} & 0 & \tfrac{1}{2} & 0 & 0 & \tfrac{1}{2} \end{pmatrix}, \nonumber \\
\vec{e}_{38} &=& \begin{pmatrix} 0 & \tfrac{1}{2} & 0 & \tfrac{1}{2} & 0 & \tfrac{1}{2} & \tfrac{1}{2} & 0 & \tfrac{1}{2} & 0 \end{pmatrix}, \nonumber \\
\vec{e}_{39} &=& \begin{pmatrix} 0 & \tfrac{1}{2} & 0 & \tfrac{1}{2} & \tfrac{1}{2} & 0 & 0 & \tfrac{1}{2} & \tfrac{1}{2} & 0 \end{pmatrix}, \nonumber 
\end{eqnarray}
\begin{eqnarray}
\vec{e}_{40} &=& \begin{pmatrix} 0 & \tfrac{1}{2} & 0 & \tfrac{1}{2} & \tfrac{1}{2} & 0 & \tfrac{1}{2} & 0 & 0 & \tfrac{1}{2} \end{pmatrix}, \nonumber \\
\vec{e}_{41} &=& \begin{pmatrix} 0 & \tfrac{1}{2} & \tfrac{1}{2} & 0 & \tfrac{1}{2} & 0 & 0 & \tfrac{1}{2} & 0 & \tfrac{1}{2} \end{pmatrix}, \nonumber \\
\vec{e}_{42} &=& \begin{pmatrix} 0 & \tfrac{1}{2} & \tfrac{1}{2} & 0 & 0 & \tfrac{1}{2} & 0 & \tfrac{1}{2} & \tfrac{1}{2} & 0 \end{pmatrix}, \nonumber \\
\vec{e}_{43} &=& \begin{pmatrix} \tfrac{1}{2} & 0 & 0 & \tfrac{1}{2} & 0 & \tfrac{1}{2} & 0 & \tfrac{1}{2} & \tfrac{1}{2} & 0 \end{pmatrix}, \nonumber \\
\vec{e}_{44} &=& \begin{pmatrix} \tfrac{1}{2} & 0 & 0 & \tfrac{1}{2} & 0 & \tfrac{1}{2} & \tfrac{1}{2} & 0 & 0 & \tfrac{1}{2} \end{pmatrix}, \nonumber \\
\vec{e}_{45} &=& \begin{pmatrix} \tfrac{1}{2} & 0 & \tfrac{1}{2} & 0 & 0 & \tfrac{1}{2} & 0 & \tfrac{1}{2} & 0 & \tfrac{1}{2} \end{pmatrix}, \nonumber \\
\vec{e}_{46} &=& \begin{pmatrix} \tfrac{1}{2} & 0 & 0 & \tfrac{1}{2} & \tfrac{1}{2} & 0 & 0 & \tfrac{1}{2} & 0 & \tfrac{1}{2} \end{pmatrix}, \nonumber \\
\vec{e}_{47} &=&  \begin{pmatrix} 0 & \tfrac{1}{2} & \tfrac{1}{2} & 0 & 0 & \tfrac{1}{2} & \tfrac{1}{2} & 0 & 0 & \tfrac{1}{2} \end{pmatrix}, \nonumber \\
\vec{e}_{48} &=& \begin{pmatrix} 0 & \tfrac{1}{2} & 0 & \tfrac{1}{2} & 0 & \tfrac{1}{2} & 0 & \tfrac{1}{2} & 0 & \tfrac{1}{2} \end{pmatrix}.
\end{eqnarray}
Now, all vertices of the non-disturbance polytope are connected among each other by scenario-preserving IMMs, 
that is -- see the diagram in Eq.(\ref{eq:diag}) -- for every $\alpha, \beta=1,\ldots, 48$ there is a matrix, let us denote it
as $\mathcal{W}(\alpha \rightarrow \beta)$, satisfying Eqs.(\ref{eq:block}), (\ref{eq:normw}) and (\ref{eq: general constraints}), such that
\begin{equation}
  \mathcal{T} \mathcal{W}(\alpha\rightarrow \beta) \mathcal{M} \vec{e}_\alpha +  \mathcal{T} \mathcal{W}(\alpha\rightarrow \beta) \vec{V} = \vec{e}_\beta;
\end{equation}
for example a possible choice of $\mathcal{W}(1\rightarrow 48)$ is 
(denoting as $W_{ij}$ is $20\times20$ matrix elements)
\begin{align}
&W_{1,1}=W_{1,3}=1,\,\, W_{1,2}=W_{2,2}=W_{2,4}=W_{3,4}=\tfrac12, \nonumber\\
&W_{5,5}=1,\quad W_{5,6}=W_{5,7}=W_{6,6}=W_{7,7}=W_{8,8}=\tfrac12, \nonumber\\
&W_{9,9}= W_{9,10}=1,\,\, W_{9,11}=W_{10,12}=W_{11,11}=W_{11,12}=\tfrac12, \nonumber\\
&W_{13,13}= W_{13,14}=W_{13,15}=1,\,\, W_{13,16}= W_{15,16}=\tfrac12, \nonumber\\
&W_{17,17}=W_{17,18}=W_{17,19}=1,\,\, 
W_{18,20}=W_{19,20}=\tfrac12, \nonumber\\
 &\text{all other elements} \,\, = 0.
\end{align}
As a consequence, for any vector of probabilities $\vec{p}$ 
within the non-disturbance polytope, and thus any quantum vector $\vec{q}$, and any vertex of the polytope, there is an IMM map $\mathcal{W}$ 
connecting the two.
In fact, $\vec{p}$ can be written as a convex combination of the vertices of the non-disturbance polytope,
\begin{equation}
\vec{p} = \sum_{\beta=1}^{48} \lambda_{\beta} \vec{e}_{\beta}, \quad \lambda_{\beta} \geq 0\,\,
\sum_{\alpha=1}^{48} \lambda_{\alpha} = 1,
\end{equation}
but then using the maps above with respect to the same $\vec{e}_{\alpha}$, we get
\begin{equation}
\vec{p} = \mathcal{T}\mathcal{W}\mathcal{M} \vec{e}_{\alpha} + \mathcal{T}\mathcal{W}\vec{V} , \quad \mathcal{W} = \sum_{\beta=1}^{48} \lambda_{\beta} W(\alpha \rightarrow \beta).
\end{equation}
Indeed, this means in particular that, as anticipated, any (quantum) probability vector within the non-disturbance polytope
can be reached via a scenario-preserving IMM from any vertex of the noncontextuality polytope.

We conclude that measurement invasiveness is a general independent mechanism to account for contextuality: 
any contextual model can be linked to a classical one solely by means of a scenario-preserving IMM. 
Even more, since any vector of contextual probabilities will be generally reachable in different ways by applying IMMs to classical probabilities, 
it is natural to ask which way is the most convenient one: answering to this question in the next section will directly lead us to introduce a distinct quantifier of contextuality.

\section{Quantifying contextuality through invasiveness}\label{sec:qct}
As the main application of our approach, we now show that scenario-preserving IMMs
can be used to define a contextuality quantifier, which can be understood as the cost of the invasiveness
required to reproduce contextual probabilities starting from classical ones.

\subsection{Invasiveness cost}
To define our quantifier of contextuality, it is useful to introduce explicitly, for
every vector of independent quantum probabilities $\vec{q}$,
the corresponding set $\mathscr{B}(\vec{q})$ formed by the scenario-preserving IMMs $\mathcal{W}$
such that Eq.(\ref{eq:convexred}) holds,
i.e. that yield a description of the quantum model via invasive measurements on a classical model.
Firstly, we note that there cannot exist a single IMM matrix $\mathcal{W}$ allowing to reproduce all quantum probabilities,
i.e. that is common to all $\mathscr{B}(\vec{q})$. 
The set of all classical models $\vec{C}$
is in fact contained within the set of all quantum models $\vec{Q}$ associated with the same $\mathfrak{F}$ \footnote{This is the case since we are not
setting any restriction on the (finite) dimension of the Hilbert space where states and measurements defining $\vec{Q}$
are settled.} and $\mathcal{W}$, being a stochastic matrix, is non-expansive --  
subset relations are preserved by $\mathcal{T}$.
On the other hand, we expect that for general scenarios every $\vec{q}$ can be reproduced if one allows for different IMMs, meaning that $\mathscr{B}(\vec{q})$ is a non-empty set
for every $\vec{q}$, as shown in the previous section 
for $\mathfrak{F}_{KBCS}$.

Given a quantum probability vector $\vec{q}$, the contextuality quantifier ${\sf IC}(\vec{q})$, which we dub 
\emph{invasiveness cost}, 
is defined as the minimum cost of a scenario-preserving IMM $\mathcal{W}$ 
that allows for the reproduction of the quantum model from some classical model,
where the cost of $\mathcal{W}$ is understood as its difference from the identity map $\mathcal{I}$, i.e., the non-invasive map:
\begin{eqnarray}\label{eq:icqn}
{\sf IC}(\vec{q})&=&\min_{\mathcal{W} \in \mathscr{B}(\vec{q})} \| \mathcal{W} -\mathcal{I}\|_2.
\end{eqnarray}
Here, $\|\cdot\|_2$ denotes the Frobenius norm, which we use to quantify the distinguishability between IMMs; indeed, ${\sf IC}(\vec{q}) = 0$ if and only if $\vec{q} \in \mathcal{N}_{\mathfrak{F}}$.

To explain more explicitly how to evaluate ${\sf IC}(\vec{q})$, let us go back to the KCBS scenario. Here, 
the corresponding scenario-preserving conditions in Eq.(\ref{eq: general constraints}),
along with Eqs.(\ref{eq:block}) and (\ref{eq:normw}), leave us with 30 independent matrix elements $W^{(i)}_{j,k}$,
which can be expressed in terms of elements of the couple $(\mathcal{Z}, \vec{v})$ via Eq.(\ref{eq:constred}).
Thus, denoting by $\vec{y}$ the 30-dimensional vector with independent $\mathcal{Z}_{j,k}$ and $v_i$,
the evaluation of ${\sf IC}(\vec{q})$ can be formulated as the following problem:
\begin{equation}\label{eq:noconv}
\begin{array}{ll}
\texttt{Find} & \vec{y} \in \mathbb{R}^{30} \\
\texttt{minimizing} &  \| \mathcal{W}(\vec{y}) -\mathcal{I}\|_2 \\
\texttt{subject to} & g_{\alpha}(\vec{y}) \leq 0 \quad \alpha=1,\ldots,80 \\
\texttt{and} & h_{\beta}(\vec{y}) \leq 0 \quad \beta=1,\ldots,16,
\end{array}
\end{equation}
where the functions $g_{\alpha}(\vec{y})$ express
the positivity of the matrix elements of the IMM $\mathcal{W}$ -- each of which is a linear combination of the elements of $\vec{y}$ fixed by Eqs.(\ref{eq:block}), (\ref{eq:normw}) and (\ref{eq: general constraints})--
while the functions $h_{\beta}(\vec{y})$ set the requirement that the counter-image of $\vec{q}$ belongs to the noncontextuality polytope
according to Eq.(\ref{eq:convexred}).
Even though the constraint functions $g_{\alpha}(\vec{y})$ and the cost function to minimize are convex,
the problem above is not a convex optimization problem
due to the non-convextity of $h_{\beta}(\vec{y})$.

\subsubsection*{A case study: quantum probabilities from experiment}
{To illustrate constructively how the proposed measure of invasiveness cost of contextuality can be evaluated from experimental data,
let us first take into account a fixed quantum probability
$\vec{q} $. Namely, consider the quantum probabilities measured experimentally in \cite{Lapkiewicz2011} on a qutrit consisting of three polarization and spatial modes of a single photon \footnote{The values are those of Table 1.(a) in \cite{Lapkiewicz2011}, taking into account that their values $A_i = \pm 1$ corresponds to ours $A_i = \mp 1$, and we are here identifying $A_1'$ wƒith $A_1$, i.e., we are not dealing with deviations from the ideal case.}:
\begin{equation}\label{eq:qexp}
\vec{q} = \left(0,0.432,0,0.473,0,0.426,0,0.439,0,0.469\right),
\end{equation}
which yields maximal violation of the KCBS inequality.
Applying the affine map $(\mathcal{M}, \vec{V})$, we get the whole vector of quantum probabilities -- see the diagram in Eq.(\ref{eq:diag}) --
\begin{eqnarray}
&&\vec{Q} = \mathcal{M} \vec{q} + \vec{V} \nonumber\\
&&= \left(0, 0.432, 0.473, 0.095, 0, 0.473, 0.426, 0.101, 0, 0.426, \right. \nonumber\\
&&\left. 0.439, 0.135, 0, 0.439, 0.469, 0.092, 0, 0.469, 0.432, 0.099\right). \nonumber
\end{eqnarray}
Now, consider the following matrix
\begin{equation}
 \mathcal{W}_p=\Theta_b \oplus \Theta_a \oplus \Theta_c \oplus \Theta_c \oplus \Theta_c, 
\end{equation}
with 
\begin{align}
   & \Theta_a=
    \begin{pmatrix}
        0&&1&&0&&0\\
        1&&0&&0&&0\\
        0&&0&&0&&1\\
        0&&0&&1&&0
    \end{pmatrix}
\quad 
    \Theta_b=
    \begin{pmatrix}
        0&&0&&1&&0\\
        0&&0&&0&&1\\
        1&&0&&0&&0\\
        0&&1&&0&&0
    \end{pmatrix}
\nonumber\\
&    \Theta_c=
    \begin{pmatrix}
        0&&0&&0&&1\\
        0&&0&&1&&0\\
        0&&1&&0&&0\\
        1&&0&&0&&0
    \end{pmatrix},\label{eq:perm1}
\end{align}
and the vector
\begin{eqnarray}
&&\vec{C} =
\left(0.473, 0.095, 0, 0.432, 0.473, 0, 0.101, 0.426,  0.135,\right. \nonumber\\
&&\left. 0.439, 0.426, 0, 0.092, 0.469, 0.439, 0, 0.099, 0.432, 0.469, 0\right)  \nonumber\\
\end{eqnarray}
such that $\vec{Q} = \mathcal{W}_p \vec{C}$.
As can be readily checked, $\mathcal{W}_p$ is a scenario-preserving IMM, i.e., it satisfies Eqs.(\ref{eq:block}), (\ref{eq:normw}) and (\ref{eq:const2}),
and $\vec{C}$ is a classical vector of probabilities belonging to the KCBS polytope, i.e., it satisfies the inequalities in Eq.(\ref{eq:polyt})
fixed by Eq.(\ref{eq:kcbsineq16}). Thus, the given quantum probabilities can be reproduced by a classical
probabilities and a linear transformation consistently accounting for invasiveness.
Even more, $\mathcal{W}_p$ is made of permutation matrices, meaning that it simply describes
swaps of the couples of measurement outcomes within each context, e.g. $\Theta_a$ describes the swaps
$$
(+1,+1) \longleftrightarrow (+1,-1) \qquad (-1,+1) \longleftrightarrow (-1,-1).
$$
Thus, for the case at hand context-dependent permutations are enough to map classical probabilities into the quantum ones maximally violating the KCBS inequality.

Of course, $\mathcal{W}_p$ is not the only possible invasive explanation connecting a vector of classical probabilities to the given quantum probabilities, 
and the invasiveness cost ${\sf IC}(\vec{q})$ precisely assesses which of these explanations is the least invasive.
Applying the procedure in Eq.(\ref{eq:noconv}) to the quantum probabilities in Eq.(\ref{eq:qexp}), we find ${\sf IC}(\vec{q}) \approx 1.851$
(while $\|W_p - -\mathcal{I}\|_2 = 2\sqrt{10}$), and the optimal IMM given by Eq.(\ref{eq:block})
with the context-independent choice $W^{(1)} = \ldots = W^{(5)}$, with
\begin{equation}\label{eq:w1cs}
W^{(1)} = \left(
\begin{array}{cccc}
 0.761 & 0 & 0 & 0 \\
 0 & 0.761 & 0.239 & 0.239 \\
 0 & 0.239 & 0.761 & 0.239\\
 0.239 & 0 & 0 & 0.522  \\
\end{array}
\right).
\end{equation}
We note that this is not a bistochastic matrix, so that it cannot be written as a convex combination of permutation matrices:
minimum invasiveness cost is not associated with swaps of the measurement outcomes, even if we allow for randomness in such swaps.
The classical vector mapped to $\vec{Q}$ by the optimal IMM is 
\begin{eqnarray}
&&\vec{C} =
\left(0, 0.37, 0.448, 0.182, 0, 0.448, 0.358,  0.194, 0, 0.358,\right. \nonumber\\
&&\left. 0.383, 0.259, 0, 0.383, 0.441, 0.176, 0, 44, 0.37, 0.19\right), \nonumber
\end{eqnarray}
which saturates the KCBS inequality.}

\subsubsection*{Dependence on mixture and superposition of quantum states}
The procedure described above in detail for a single vector of quantum probabilities can be repeated for generic quantum states.
In particular, we looked at the invasiveness cost ${\sf IC}(\vec{q})$ for a family of quantum probabilities $\vec{q}$,
which includes those arising from the state $\ketbra{1}{1}$ leading to the maximal violation in Eq.(\ref{eq:max}), 
as well as other pure states 
and their mixtures with the maximally mixed state.
Explicitly, we considered the states
\begin{equation}\label{eq:rhoaa}
\rho = (1-\lambda) \frac{\mathbbm{1}}{3} + \lambda\ket{\psi_a}\bra{\psi_a}, \quad \ket{\psi_a}=a \ket{1} + \sqrt{1-a^2}\ket{2},
\end{equation}
with corresponding vector of quantum probabilities -- see Eqs.(\ref{eq:pix}) and (\ref{eq:qx}) --
\begin{eqnarray}
    && \hspace{-0.5cm} (\vec{q})_{2i-1} = 0 \label{eq:qaa}\\
 &&   \hspace{-0.5cm} (\vec{q})_{2i} = \frac{1-\lambda}{3}+\lambda\left(\frac{a}{\sqrt[4]{5}}+\frac{\sqrt{(1-a^2)(\sqrt{5}-1)}}{\sqrt[4]{5}} \cos \varphi_i\right)^2.  \nonumber
\end{eqnarray}
\begin{figure}[t]
    \centering
    \includegraphics[width=0.41\textwidth]{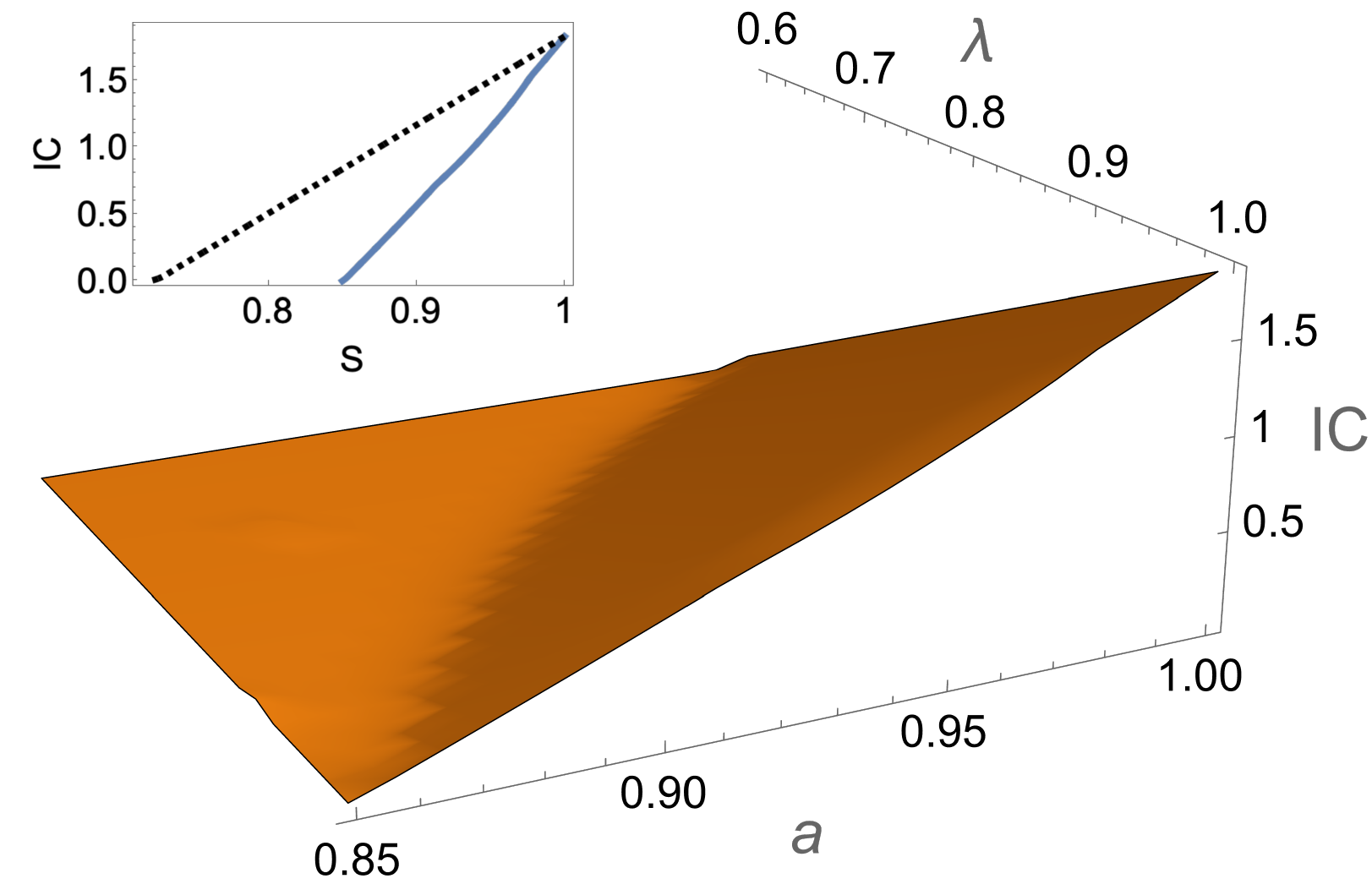}
  \caption{Invasiveness cost ${\sf IC}(\vec{q})$ as a function of the mixing parameter $\lambda$ and superposition parameter $a$,
for the family of quantum states $\rho$ in Eq.(\ref{eq:rhoaa}) leading to the probabilities in Eq.(\ref{eq:qaa}).
(Inset) Invasiveness cost ${\sf IC}(\vec{q})$ as a function of, respectively, $S=a$ or $S=\lambda$,
for the quantum states $\ket{\psi_a}$ (solid line) and $\lambda \ket{\psi_1}\bra{\psi_1}+(1-\lambda) \ket{\psi_0}\bra{\psi_0}$ (dashed line).}
  \label{fig:ic}
\end{figure}
In Fig.\ref{fig:ic}, we report the invasiveness cost ${\sf IC}(\vec{q})$ evaluated via the procedure in Eq.(\ref{eq:noconv}), 
as a function of the mixing and superposition parameters.
As expected, ${\sf IC}(\vec{q})$ is equal to zero for all probabilities arising from states that satisfy the KCBS inequality, 
while it is maximal for the state $\ketbra{1}{1}$. 
In particular, ${\sf IC}(\vec{q})$ decreases monotonically with increasing the mixture and/or superposition of $\ketbra{1}{1}$ with the other states,
with a different decreasing rate depending on whether a mixture or a superposition is considered -- see also the inset.

The whole set of data, including the optimal IMMs $\mathcal{W}$ achieving the minimum in Eq.(\ref{eq:noconv}), can be found at \cite{zeno}.
In particular, one can see that the optimal IMMs typically have the same symmetric structure, $W^{(1)} = \ldots = W^{(5)}$, 
shown by the case study leading to Eq.(\ref{eq:w1cs}). This reflects the symmetry under the relabeling of the contexts $i \mapsto i+1$ for any $i$
possessed by both the scenario-preservation conditions in Eq.(\ref{eq: general constraints}) and the connection 
with the noncontextuality polytope expressed by Eq.(\ref{eq:convexred}).

\subsection{Comparison with other contextuality quantifiers}

Compared to other quantifiers of contextuality \cite{Kleinmann2011,Grudka2014,Abramsky2017,Amaral2017,Amaral2018,Kujala2019,Horodecki2023}, 
invasiveness cost targets the stochastic maps that connect noncontextual to contextual probabilities,
rather than directly comparing the two types of probabilities or assessing the amount of deterministic assignments needed to reproduce
contextual predictions.
In other terms, invasiveness cost singles out how much intervention is required to reproduce contextual probabilities from classical ones,
rather than directly addressing how different the former are from the latter.
Despite such an operational difference, it is a-priori not obvious to what extent different quantifiers 
of contextuality result in different relations among contextual probabilities.
Here, we provide an explicit comparison of the invasiveness cost ${\sf IC}(\vec{q})$ with 
contextual fraction \cite{Abramsky2011,Abramsky2017}, 
focusing on quantum probabilities of $\mathfrak{F}_{KCBS}$, for $\mathcal{H}=\mathbbm{C}^3$ and the 
projective measurements fixed by the self-adjoint operators $\widehat{A}_i$ in Eq.(\ref{eq:sta}). 

Starting from the decomposition of a model into a mixture of any noncontextual model and a further (contextual) model,
contextual fraction was introduced in \cite{Abramsky2011} as the minimum possible value of the mixing parameter of the latter:
in our notation,
\begin{equation}\label{eq:cfe0}
    {\sf CF}(\vec{q}) = \min_{0 \leq \lambda \leq 1}  \left\{\lambda \,\, \mbox{s.t.}\,\,\vec{q} = \lambda \vec{q}' + (1-\lambda) \vec{c}\right\},
\end{equation}
where $\vec{q}'$ is any vector of quantum probabilities and $\vec{c}$ is any vector within the noncontextuality polytope.
\begin{figure}
\centering
\includegraphics[width=0.47\textwidth]{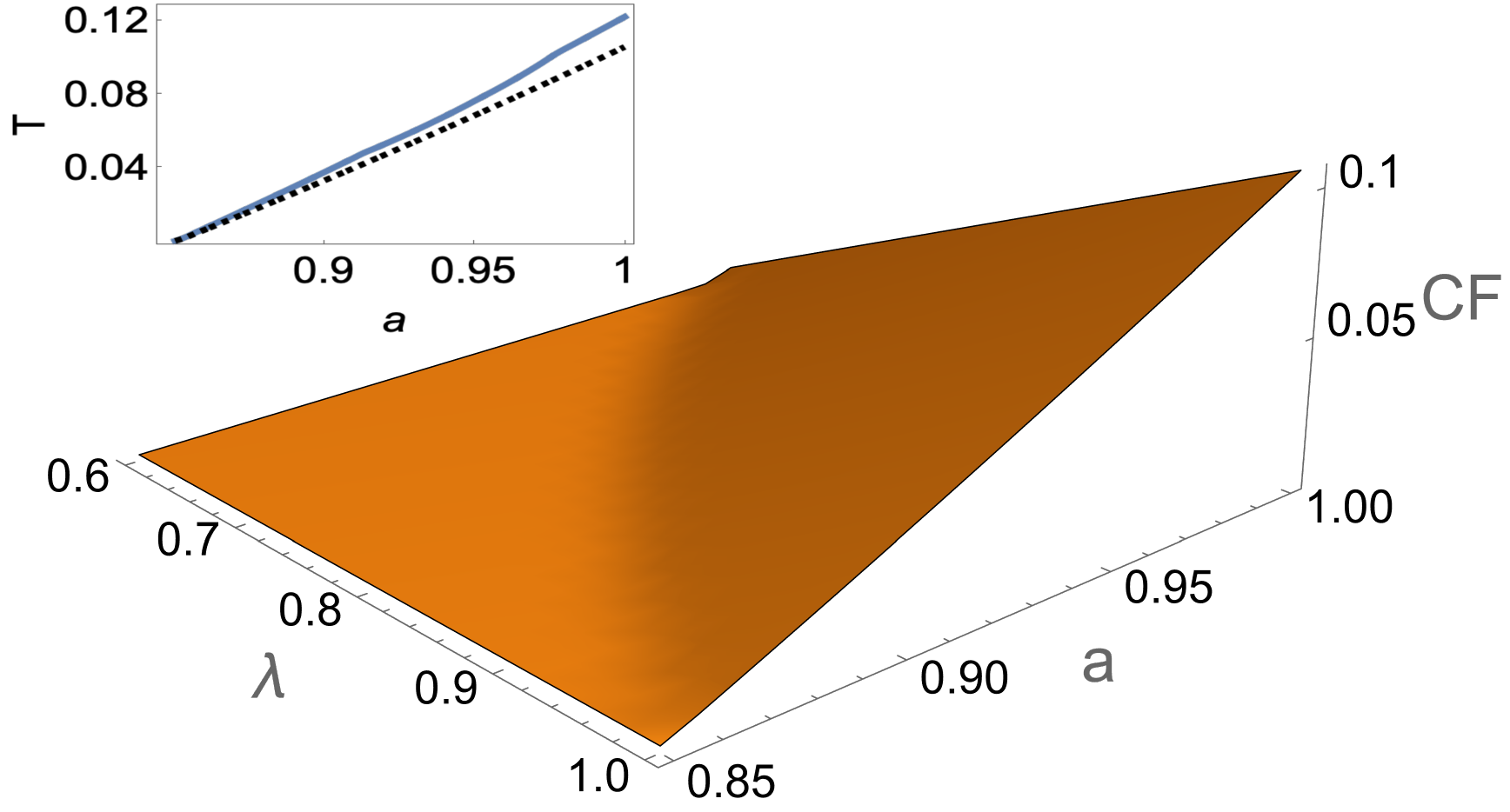}
\caption{Contextual fraction ${\sf CF}(\vec{q})$ as a function of $\lambda$ and $a$,
for the same family of quantum states as those in Fig.\ref{fig:ic}.
(Inset) Invasiveness cost ${\sf T} = {\sf IC}(\vec{q})$ (solid line)
and contextual fraction ${\sf T} = {\sf CF}(\vec{q})$ (dashed line) as a function of $a$
for the probabilities given by the quantum state $\ket{\psi_a}$;
${\sf IC}(\vec{q})$ is divided by 15 to have a comparable scale between the two quantities.}\label{fig:cfe}
\end{figure}
In \cite{Abramsky2017}, it was then shown that the contextual fraction of a contextual model 
can be equivalently expressed as the (normalized) violation of a noncontextual inequality.
Recalling that for $\mathcal{H}=\mathbbm{C}^3$ and the projective measurements in Eq.(\ref{eq:sta}) 
only the first of the 16 inequalities defining the polytope $\mathcal{N}_{KCBS}$ can be violated,
the contextual fraction can thus be written as
\begin{equation}\label{eq:cfe}
    {\sf CF}(\vec{q}) = \frac{\vec{q} \cdot \vec{f}_1-b_1}{\|f_1\| - b_1},
\end{equation}
with $\vec{f}_1$ and $b_1$ given by Eq.(\ref{eq:kcbsineq16}).
We note that for the KCBS scenario (more generally, for any $n-$cycle scenario), this is also proportional to 
the generalized robustness of contextuality \cite{Meng2016}, that is, the minimum amount of any (classical or quantum) vector of probabilities 
that has to be mixed with the given probability vector to make it contextual.

In Fig. \ref{fig:cfe}, we report the contextual fraction ${\sf CF}(\vec{q})$ 
for the same family of states as those considered for invasiveness cost ${\sf IC}(\vec{q})$
in Fig. \ref{fig:ic}. We can observe the same kind of behavior, in the sense that both quantifiers
order contextual probabilities in the same way. Indeed, such a connection is expected since both quantifiers are simply
assessing how much $\vec{q}$ is violating the only non-trivial KCBS inequality. 
On the quantitative side, besides for an overall factor, 
the difference among the two quantifiers varies within the considered set of states -- see the inset.
We leave for future investigation the analysis of more complex situations, 
where different inequalities might have different relevance for distinct quantifiers.\\

\section{Conclusions}\label{sec:con}
In this work we introduced a general framework to analyze quantum contextuality in terms of measurement invasiveness, formalized as stochastic linear maps acting on classical (noncontextual) probabilities. Within the marginal-scenario approach to contextuality, we showed how IMMs can be consistently defined, to preserve the compatibility structure of a given scenario, while allowing classical probabilities inside the noncontextuality polytope to be transformed into contextual ones outside it.
Our approach yields a clear separation between classicality of the underlying statistics and invasiveness of the measurement procedure, 
allowing one to isolate invasiveness as an independent source of contextuality.\\
Building on this framework, we further introduced the invasiveness cost as a quantitative measure of contextuality, 
defined as the minimum Frobenius distance from the identity map of an IMM required to reproduce a given probability distribution. Unlike standard quantifiers, which directly compare probability distributions, invasiveness cost targets the transformations connecting classical and contextual models, quantifying how much one must perturb an ideal non-invasive measurement in order to reproduce a contextual behavior starting from classical statistics. \\
As a case study, we fully characterized the admissible IMMs for the KCBS scenario, 
also showing that all quantum KCBS probabilities can be obtained from any vertex of the noncontextuality polytope via a proper IMM.
We further fully characterized the invasiveness cost for a vector of quantum probabilities from experiment \cite{Lapkiewicz2011},
providing a clear operational meaning for the corresponding optimal maps,
as well as for a family of quantum states, involving mixtures and superpositions with the state leading to a maximal violation of the KCBS inequality.\\
Several directions for future research naturally emerge from our results. First, while we focused on jointly measurable observables, the formalism of invasive maps extends naturally to sequential-measurement scenarios \cite{Fritz2010,Hoffmann2018,Moreira2019,Milz2020,Vitagliano2023,Richter2024}, where invasiveness and memory effects play a central role and a clear distinction among them is still missing. Second, applying our framework to Bell scenarios will allow us to investigate nonlocal correlations in terms of invasive, nonlocal, maps, looking for a unified approach for contextuality and nonlocality at the level of probability transformations.
More broadly, it will be important to study the relation between our approach and different frameworks for contextuality, including measurement, preparation, and transformation contextuality \cite{Spekkens2005}.
Invasive maps might also be useful for connecting contextuality with approaches to the description of quantum systems via classical tools, 
such as the classical simulation schemes based on probabilistic updates of probability distributions on a finite classical phase space  \cite{Zurel2020},
where quantum computational power is associated with more general resources than contextuality alone.

\section*{Acknowledgments}
The authors acknowledge support from MUR and
Next Generation EU via the
PRIN 2022 Project “Quantum Reservoir Computing
(QuReCo)” (contract n. 2022FEXLYB) and the NQSTI-Spoke1-BaC
project QSynKrono (contract n. PE00000023-QuSynKrono).

\end{document}